\documentclass[12pt]{article}                  %% LaTeX 2e
\usepackage{osajnl}
\usepackage{hyperref}

\begin{document}

\title{A high-yield single photon source using gated spontaneous parametric down conversion}

\author{Shigeki Takeuchi}
%% for REVTeX4, each author name can be set in a separate \author{} field

\address{Japan Science and Technology Corporation-
Precursory Research for Embryonic Science and Technology Project,\\
Research Institute for Electronic Science, Hokkaido University, Kita-12 Nishi-6,Kita-ku, Sapporo, Hokkaido 060-0812 Japan}

\author{Ryo Okamoto, and Keiji Sasaki}
%% for REVTeX4, each author name can be set in a separate \author{} field

\address{Research Institute for Electronic Science, Hokkaido University, Kita-12 Nishi-6, Kita-ku, Sapporo, Hokkaido 060-0812 Japan}

\begin{abstract}The construction of a single photon source using gated parametric fluorescence is reported with the measurement results of the photon number distribution. A beamlike twin-photon method is used in order to achieve high collection efficiency. The estimated probability $P(1)$ to find a single photon in a collimated output pulse is 26.5 \% at a repetition rate of 10 kHz when the effective quantum efficiency of 27.4 \% in the detection setup is compensated.
\end{abstract}

%\ocis{000.0000, 999.9999.}% REPLACE WITH CORRECT OCIS CODES FOR YOUR ARTICLE
                          % NOTE: \ocis{} IS ALIASED TO \pacs{} BUT MUST
                          % FORMAT THE TERMS CORRECTLY FOR EACH JOURNAL
\maketitle %% NULL FUNCTION WITH LATEX 2e

\section{Introduction}

One of the most important applications of the single photon source is quantum cryptography\cite{BB84}. Up to now, most of the experiments of quantum cryptography have been using weak coherent light as photon sources \cite{Wak00}. In those experiments, the average photon number was kept as small as 0.1 in order to keep the probability of generating two photons in one pulse very low. However, this meant that 90 \% of the pulses were vacuum states and could not be used to send any information. Therefore, single photon sources which can output single photon states with a high probability while having only small percentages of two photon states are very important for quantum cryptography. Such single photon sources are also important for quantum information processing. 

There has been extensive research to produce single photon pulses using a single light emitter. Experimental demonstrations have been performed using single molecules\cite{Mar96,Tre02}, single color centers in diamond \cite{Kur00} and single quantum dots \cite{Sant01} optically pumped by short pulse lasers.  To realize electrically driven solid state single photon sources, the pioneering work using turnstile devices at very low temperatures \cite{Kim99} has recently been followed by a demonstration of operation at 4.2K \cite{Yua02}. In those methods, however, photons are emitted into all directions and it is difficult to collect the photons with high efficiency. This means that effectively the percentage of single photon pulses will be small and that of vacuum states will be large. To the best of our knowledge, the highest probability $P(1)$ to find a single photon at an output port where single photons are collimated was less than 10\% in such devices. Recently, Pelton et. al. reported an excellent experimental result using a quantum dot embedded in a semiconductor micropost microcavity, where the efficiency of emitting a single mode traveling wave inside the cavity was 38\% \cite{Pel02}. However, the overall detection efficiency was only 3\% due to the technical difficulties of collecting the single photons and filtering them out from background photons. 

The generation of single photon states using spontaneous parametric down-conversion (SPDC) is an alternative approach, which was first mentioned by Hong and Mandel \cite{Hon86}. 
It was theoretically followed by Yuen \cite{Yue86}, who studied the parametric processes with measurement feedback for the generation of near-photon-number-eigenstates. 
The idea was experimentally followed by several groups\cite{Tas86,Rar87,Mer90}.  However, these experiments focused on the observation of sub-Poissonian photon statistics, and not the demonstration of a true single-photon source. Therefore, they just measured the correlation function $g^{(2)}( \tau )$ and determined the Fano factor $F = (\Delta n)^2 / <n>$. 
Both properties characterize the average noise level of the photon current, but the photon number distribution cannot be derived from only these parameters\cite{comment}.

In this paper, we present a single photon source based on gated spontaneous parametric down-conversion (G-SPDC). In contrast to earlier work\cite{Tas86}, our G-SPDC source is run in a new regime in which a single-photon is emitted in each time window with high probability and a well defined repetition rate. Furthermore, the quality of our single photon source is measured by characterizing the photon number distribution of its output.

\section{Experimental Setup and Data Analysis}

A schematic of our single photon source is shown in Figure 1. When a continuous wave ( CW ) ultra violet ( UV ) pump beam at 351.1 nm is incident to a nonlinear crystal fulfilling phase matching condition, two pairs of single visible photons, control photons and signal photons, are created at the same time through the process of spontaneous parametric down conversion (SPDC)\cite{Man95}. We adjusted the experimental parameters to have a sufficient number of photon pairs created in a pulse with 100 $\mu$s pulse width. In our experiment, an average of eight control photons is detected within the pulse width. Following the first photon detection of a control photon in the pulse, the gate timing controller opens the shutter for a very short time. This means that only the first signal photon in the pulse can pass through the shutter and be emitted. After the emission, the shutter is closed to block the other photons in the pulse. The repetition of this procedure every 100 $\mu$s results in a repetition rate of 10 kHz in our experiment. 
We have reported that parametric fluorescence can be emitted into two small spots and that the rate of coincident counts to singles counts can be larger than 0.8 \cite{Tak01}. In this experiment, we adopted this condition to increase the collection efficiency of photons, and also focused the pump laser beam softly with a convex lens to obtain a high yield of parametric fluorescence \cite{Mon98}.

During the time it takes to control the shutter according to the detection signal of the control photon, which is 150 ns in our experiment, the signal photon should be stored with minimal losses.  For this purpose, we adopted an optical delay line using a multi mode optical fiber with a transmittance of 50 \%, including the losses that occurs at the input/output fiber couplers and at the UV cut filters inserted in front of the fiber coupler.

We introduced a fast optical shutter with high transmission in this experiment. 
For fast optical switching, optical shutters using electro optic modulators ( EOMs ) are generally used. However, such EOM modulators are polarization dependent and not appropriate for the signal photon output from the multi-mode fiber delay line in which the polarization has been randomized. Therefore, we used the optical setup for the shutter shown in Figure 1. To begin with, the first polarizing beam splitter ( PBS ) transforms the signal photon into a superposition of the states in each of the paths according to its polarization.
Then the electro optic modulators ( EOM ) ( LM0202, Linos ) in both paths rotate the polarization by 90 degrees only when the shutter is controlled to be closed. Finally, the two components are combined again by the second PBS. We used two EOMs for each path in order to minimize the shutter opening time and to avoid the restriction for repetition time given by the EOM drivers ( LIV-8, Linos ). In the experiment, the shutter opening time was set to 50ns with a transmittance of 83 \%. With this setup we achieved an overall effective transmittance of 41 \% for signal photons.

In order to find the photon number distribution of the output state of our single photon source ( in multi-mode ), we measured the number of photons in each output pulse ( 100 $\mu$s ) by a  photon number analyzer consisting of a single photon counting module ( SPCM-AQ-FC, Perkin Elmer )  and a photon counter ( SR-400, Stanford Research Systems ). The real photon number distribution of the output modified by the losses caused by the limited photon detection efficiency of the photon number analyzer is as follows\cite{Hon86}.
\begin{equation}
P'(i)=\sum_{j \geq i} \frac{j!}{(j-i)! i!} \eta^i (1-\eta)^{j-i} P(j),
\label{eq:P'}
\end{equation}
where $\eta$, $P(j)$ and $P'(i)$ are the effective quantum efficiency of the analyzer, the probability of having $j$ photons in the output, and the percentage of the events where the analyzer recorded $i$ signals during the pulse duration, respectively. We estimated the photon number distribution $P(j)$ of the output state by applying equation (1) to the experimental values of $P'(i)$ for $i = 0$ to 2 and the estimated effective quantum efficiency of $\eta = 0.274 \pm 0.019$, taking into account the estimated quantum efficiency of $70 \pm 5$ \% for SPCM and experimentally measured optical losses caused by the coupling lens and a mirror ( transmittance 90.2 \% ), the filter for the stray photons ( transmittance 49.2 \% ) and the fiber coupler( transmittance 88.2 \% ). As a result, the photon detection efficiency of the photon number analyzer is estimated to be 27.4 \%. Note that we compensated the photon detection efficiency of the photon number analyzer at the output, but we did not compensate the optical losses in the experimental setup of the G-SPDC single photon source.

In the estimation of $P'(i)$, the dark count of the detector of 100 cps was also subtracted from the original data. The number of photons which are not counted during the dead time ( 50 ns )of the photon counter was also compensated under the assumption that the SPDC has Poisson distribution in multi-mode\cite{comment2}.

\section{Results and Discussions}

The photon number distribution obtained in our experiment is shown by the bars (a) in Fig. 2. These $P(j)$ are estimated using Eq.\ref{eq:P'} from the raw photon counting data $P'(0)=0.9199, P'(1)=0.0794, P'(2)=0.0005$ with the measurement of $10^5$ output pulses. The probability $P(1)$ to have a single photon in a pulse is $0.265 \pm 0.02$, which is more than three times larger than the previously reported value as far as we know. The probability $P(2)$ to have two photons in a pulse was estimated to be $0.011 \pm 0.001$.  The average photon number of the output in a pulse is $0.29 \pm 0.02$. The bars (b) show the calculated photon number distribution of weak coherent light ( WCL ) with the same average photon number. The sub-Poisson characteristics of our single photon source is clearly observed since the $P(1)$ of G-SPDC is higher than that of WCL and at the same time $P(2)$ of G-SPDC is much smaller than that of WCL.

The probability $P(2)$ of two photon states indicates the security risk for quantum cryptography. The bars (c) in Fig. 2 show the photon number distribution of weak coherent light having a $P(2)$ of 0.011, which is the same as that of our source. The $P(1)$ of our source is 2 times larger than that of the weak coherent light. This result also illustrates the potential advantage of our source over weak coherent light as a single photon sources for quantum cryptography. This is our present experimental status. Next, we will analyze this photon number distribution quantitatively.

The probability $P(1)$ is restricted by the optical losses in the path of signal photons in our experimental setup. The 41 \% overall transmittance of the path suggests that the collection efficiency of the parametric fluorescence by the fiber coupler is 68 \%, which is consistent with the estimation by independently performed measurement of the absolute quantum efficiency of the detection setup \cite{Mig95}.

Ideally, we can also eliminate the probability $P(2)$ to find more than two photons in a pulse by the G-SPDC method. $P(2)$ is not ideal because the optical shutter used in our experiment is not perfect in two points. One is the non-infinitesimal shutter opening time (50 ns), and the other is the small leakage of photons through the closed optical shutter (0.1 \%). 

The following is a discussion of the possibility for applications that requires higher repetition rates. One of the applications could be QKD. Note that long shutter opening time ( 100 $\mu$s ) in the present setup would be one of the problems that prevent an immediate adoption of our source for conventional quantum key distribution (QKD) systems at 1.55$\mu$m wavelength with photon counters gated by nano-second timing pulses. In our experiment, the repetition rate of 10 kHz was limited by a photon pair emission rate of 10$^6$ per second with a pump power of 200 mW. With this photon pair emission rate, we had eight detection signals of control photons per pulse on average. We can increase the repetition rate to several tens of kHz with our present setup, however, the probability to find no control photon in a pulse becomes significant when the repetition rate is increased too much. In order to overcome this limitation, it may be possible to increase the photon emission rate up to $10^9$ per second by using a periodically poled lithium niobate ( PPLN ) crystal \cite{Sana01} and an increased pump power of 1 W.  This means that the repetition rate of several tens of MHz is possible using conventional technologies. This rate can compete with most single photon sources using single light emitters. Note also that none of the single photon sources that have been reported so far can be used to replace the weak coherent light in the conventional 1.55$\mu$m QKD system, because of available wavelength, low collection efficiencies, limited lifetimes, low operation temperatures and so on.

Our G-SPDC method has two significant advantages compared to other kinds of single photon sources. First, our method operates at room temperature and is very stable. This compares favorably with the fact that many other proposals are suffering from a short life time or require low temperature operation. The other advantage is that the wavelength of single photons can be continuously changed. The SPDC process occurs not only for the degenerate condition but also for other pairs of wavelengths when energy and momentum conservation laws are satisfied. In our case, we can easily change the wavelength of photons just by changing the angle of the BBO crystal. This tunability is a unique characteristic of the G-SPDC method, since methods using single light emitters have emission wavelength that are determined by the materials or by the device structure.

Because we adopted a multi-mode fiber optical delay line to minimize the optical losses, the single photons are emitted in multi modes in space. With regard to the bandwidth of the emitted photons, we could not measure it with our present experimental setup but we guess that it should be about 0.26 nm ( full width half maximum ). This is because we used a narrow band filter of 0.26 nm for control photons, and the corresponding signal photons which are emitted through the shutter should have a similar bandwidth due to energy conservation. In this case, the associated coherence time of the emitted photons would be 6.25 ps. Therefore, the photonic wave function is supposed to be localized in time in the whole time duration ($100 \mu$s) of the pulses.

Note that even a source which emits single photons in multi mode may be used in quantum cryptography experiments such as the one recently reported using built-in multi-mode optical fiber telecommunication network \cite{Yos02}. Generation of single photons in a single mode in space and/or in a single mode in time domain remains as a future challenge\cite{Tak02,Pit02}.

\section{Conclusions}

In conclusion, we have experimentally demonstrated the generation of single photon pulses using the G-SPDC method. The measured single photon probability was found to be $P(1)$ = 0.265 at the collimated output port at a repetition rate of 10 kHz after the effective quantum efficiency of 27.4 \% in the detection setup was compensated. The photon number distribution of the output pulses also clearly showed sub-Poissonian characteristics. Single photon sources using our method have long lifetimes at room temperatures, and can be tuned to wide range of frequencies. 

We thank K. Ushizaka for assistance with the gate control circuit; K. Tsujino for assistance; Y. Yamamoto and E. Waks for their motivating us to perform this experiment.; H. F. Hofmann, H. Ohashi, H. Fujiwara, J. Hotta for discussion. This work was supported by JST-PRESTO project, Mitsubishi Electric, International Communications Foundation and Grant-in-Aid for Scientific Research(B) 12555008 of JSPS. E-mail address, takeuchi@es.hokudai.ac.jp.

%Part of this work was supported by the program "Research and
%Development on Quantum Communication Technology" of the
%Ministry of Public Management, Home Affairs, Posts and
%Telecommunications of Japan.

%Part of this work was supported by the program "R&D on Quantum 
%Commun. Tech." of the MPHPT of Japan.

%-------------
\newpage
%-------------

{\large Figure captions}

\vspace{1cm}

Figure 1:

Schematic of the experimental setup. A signal photon and a control photon are created at the same time in a BBO crystal. The control photons are directly detected by a single photon detector while the signal photons were guided into an optical fiber delay line. Then, the signal photons were gated by the fast optical shutter. This shutter opens only when the first detection event of control photon is observed in each pulse ( time window of 100 $\mu$s ) and blocks other photons in the same pulse.

\vspace{1cm}

Figure 2:

(a) Photon number distribution of the output state. $P(1)$ and $P(2)$ shows the probability of finding one / two  photons in a output pulse (100 $\mu$s) respectively. We counted the number of detection event in each pulses and estimated a photon number distribution at the output of our source as described in the text. (b) Photon number distribution of weak coherent light with the same average photon number of (a). (c)  Photon number distribution of weak coherent light with the same $P(2)$ of (a).

%-------------
\newpage
%-------------

{\Large Figure 1}

\vspace{3cm}

%------------------ Figure 1 --------------------------------------
\begin{figure}[h]
\includegraphics{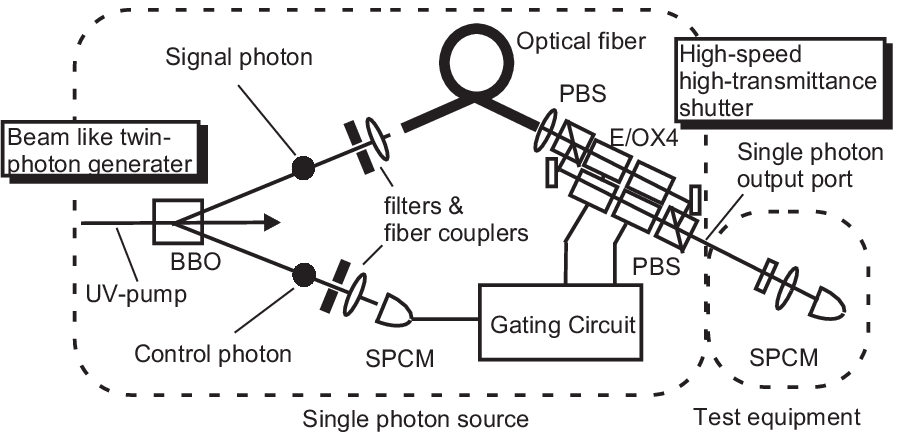}
\label{fig:Fig1} 
\end{figure}
%------------------------------------------------------------------
%

%-------------
\newpage
%-------------
%

{\Large Figure 2}

\vspace{3cm}

%------------------ Figure 2--------------------------------------
\begin{figure}[h]
\includegraphics{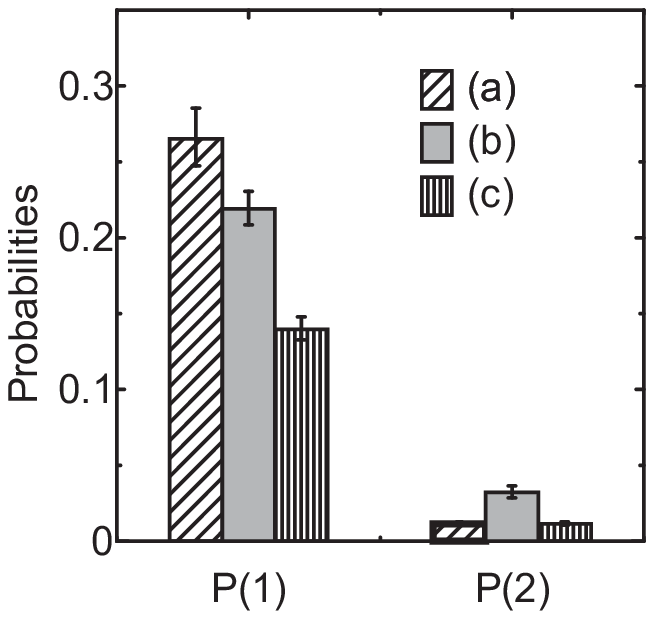}
\label{fig:Fig2} 
\end{figure}
%------------------------------------------------------------------
%


\begin{thebibliography}{99}
\bibitem{comment}
For example, a single photon source with the photon number distribution $\{P(0),P(1),P(2)\}=\{0.7, 0.3, 0\}$ is much more useful for quantum cryptography than that with $\{0.5, 0.4, 0.05\}$ , where $ P(n)$ is the probability to find n photons in a pulse. Those photon number distributions, however, give you the same Fano factor $F=0.7$.
\bibitem{comment2}
This assumption was veryfied by measuring the coincidence using Hanbury-Brown and Twiss type setup.
\end{thebibliography}
\end{document}